\documentclass[preprint,preprintnumbers, prd, floatfix, superscriptaddress,nofootinbib] {revtex4-1}
\usepackage{epsfig}
\usepackage{subfigure}
\usepackage{dcolumn}
\usepackage{bm}
\usepackage[usenames ,dvipsnames]{xcolor}
\usepackage{slashed}
\usepackage{graphicx,color}
\begin{document}
\title{Resonant $a_0(980)$ state\\
in triangle rescattering $D_s^+\to \pi^+\pi^0\eta$ decays}

\author{Yu-Kuo Hsiao}
\email{Corresponding author: yukuohsiao@gmail.com}
\affiliation{School of Physics and Information Engineering, Shanxi Normal University, Linfen 041004, China}

\author{Yao Yu}
\email{Corresponding author: yuyao@cqupt.edu.cn}
\affiliation{Chongqing University of Posts \& Telecommunications, Chongqing, 400065, China}

\author{Bai-Cian Ke}
\email{Corresponding author: baiciank@ihep.ac.cn}
\affiliation{School of Physics and Information Engineering, Shanxi Normal University, Linfen 041004, China}

\date{\today}

\begin{abstract}
We study the $D_s^+\to \pi^+(a_0(980)^0\to)\pi^0\eta$, 
$\pi^0(a_0(980)^+\to)\pi^+\eta$ decays, 
which have been recently measured by the BESIII collaboration.
We propose that $D_s^+\to \pi^{+(0)}(a_0(980)^{0(+)}\to)\pi^{0(+)}\eta$
receives the contributions from the triangle rescattering processes,
where $M^0$ and $\rho^+$ in $D_s^+\to M^0 \rho^+$,
by exchanging $\pi^{0(+)}$,
are formed as $a_0(980)^{0(+)}$ and $\pi^{+(0)}$, respectively,
with $M^0=(\eta,\eta')$. Accordingly, we calculate that 
${\cal B}(D_s^+\to a_0(980)^{0(+)}\pi^{+(0)})=
(1.7\pm 0.2\pm 0.1)\times 10^{-2}$ and
${\cal B}(D_s^+\to \pi^{+(0)}(a_0(980)^{0(+)}\to)\pi^{0(+)}\eta)
=(1.4\pm 0.1\pm 0.1)\times 10^{-2}$, being consistent with the data.
\end{abstract}

\maketitle
\section{introduction}
\begin{figure}[t!]
\includegraphics[width=2.8in]{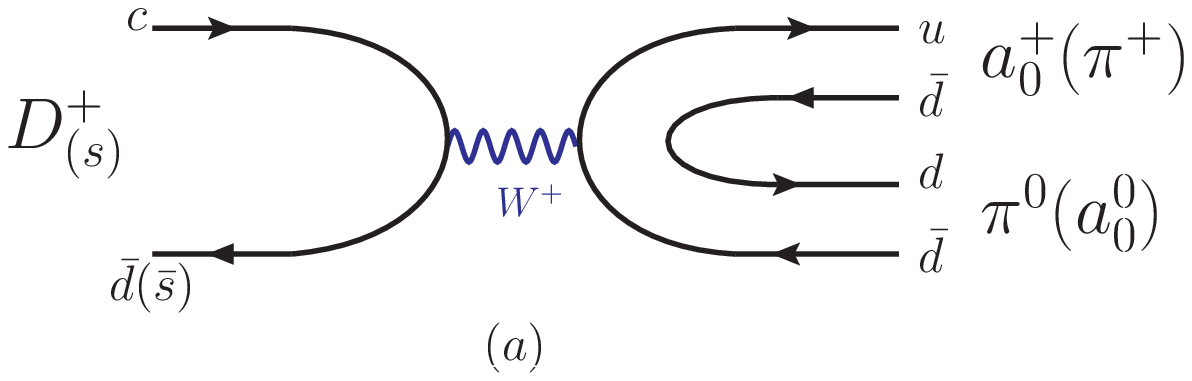}
\includegraphics[width=2.8in]{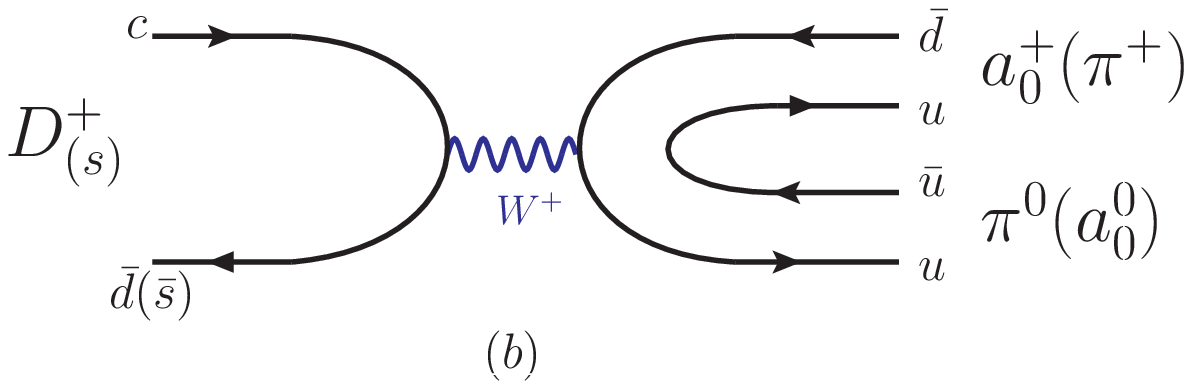}
\caption{$D^+_{(s)}\to a_0^+\pi^0$, $a_0^0\pi^+$
via the $W$-annihilation diagrams.}\label{pic_anni}
\end{figure}

Recently, the BESIII collaboration has measured 
the branching fraction of the $D_s^+$ decay that involves
one of the scalar mesons below 1~GeV, $a_0\equiv a_0(980)$,
which still has a controversial identification~\cite{Jaffe:2004ph,
Stone:2013eaa,Maiani:2004uc,Agaev:2018fvz,Wang:2009azc,Molina:2019udw}. 
Explicitly,
the branching fractions are observed as~\cite{Ablikim:2019pit}
\begin{eqnarray}\label{data1}
{\cal B}(D_s^+\to \pi^{+(0)}(a_0^{0(+)}\to)\pi^{0(+)}\eta)
=(1.46\pm 0.15\pm 0.23)\times 10^{-2}\,,
\end{eqnarray}
where the $D_s^+\to a_0^+\pi^0$, $a_0^0\pi^+$ decays
are claimed as the W-annihilation (WA) dominant  processes
observed for the first time, as depicted in Fig.~\ref{pic_anni}.
Nonetheless,
if $D_s^+\to a_0\pi$ proceeds through 
the WA $c\bar s\to W^+\to u\bar d$ decay, 
the $G$-parities of $u\bar d$ and $a_0\pi$ are 
odd and even, respectively~\cite{Cheng:2010ry,Achasov:2017edm}, 
such that $a_0\pi$ formed from $u\bar d$
violates $G$-parity conservation, indicating 
the suppressed WA process for $D_s^+\to a_0\pi$.

The same WA processes can also be applied to the $D^+$ section,
being barely allowed by the current data.
With ${\cal B}_{\text{WA}}(\eta^{(\prime)})
\equiv{\cal B}(D^+\to \pi^{+(0)}(a_0^{0(+)}\to)\pi^{0(+)}\eta^{(\prime)})$,
we obtain that
\begin{eqnarray}\label{est_1}
{\cal B}_{\text{WA}}(\eta)&\simeq&\bigg(\frac{f_D}{f_{D_s}}\bigg)^2
\bigg(\frac{|V_{cd}|}{|V_{cs}|}\bigg)^2\frac{\tau_{D}}{\tau_{D_s}}
\bigg(\frac{m_{D_s}}{m_{D}}\bigg)^3 
\times{\cal B}(D^+_s\to \pi^{+(0)}(a_0^{0(+)}\to)\pi^{0(+)}\eta)\nonumber\\
&&=(1.2\pm 0.2)\times 10^{-3}\,,
\end{eqnarray}
where $f_{D_{(s)}}$,$\tau_{D_{(s)}}$, $m_{D_{(s)}}$,
and $V_{cq}$ ($q=d,s$)
represent the decay constant, lifetime, mass for the $D^+_{(s)}$ meson,
and the Cabibbo-Kobayashi-Maskawa (CKM) matrix elements, respectively.
It has been measured that 
${\cal B}({\eta,\eta'})\equiv{\cal B}(D^+\to \pi^+\pi^0\eta,\pi^+\pi^0\eta')
=(1.4\pm 0.4,1.6\pm 0.5) \times 10^{-3}$~\cite{pdg}.
The fact of ${\cal B}(\eta)\simeq {\cal B}(\eta')$ indicates that 
$D^+\to \pi^+\pi^0\eta,\pi^+\pi^0\eta'$ have 
the same topologies except for the difference from the $\eta-\eta'$ mixing.
With ${\cal B}_\rho(\eta^{(\prime)})
\equiv {\cal B}(D^+\to \eta^{(\prime)}(\rho^+\to)\pi^+\pi^0)$ and
${\cal B}_{\text{WA}}(\eta^{(\prime)})$ that mainly
contribute to ${\cal B}(\eta^{(\prime)})$,
that is, ${\cal B}(\eta^{(\prime)})={\cal B}_\rho(\eta^{(\prime)})+{\cal B}_{\text{WA}}(\eta^{(\prime)})$,
one should have 
${\cal B}_{\rho,\text{WA}}(\eta)\simeq {\cal B}_{\rho,\text{WA}}(\eta^{\prime})$.
Nonetheless, 
due to ${\cal B}(a_0\to \pi\eta')\simeq 0$, caused by
${\cal B}(a_0\to \pi\eta+K\bar K)\simeq 100\%$~\cite{pdg},
it is estimated that
${\cal B}_{\text{WA}}(\eta')=
{\cal B}(D^+\to \pi^{+(0)} a_0^{0(+)})\times 
{\cal B}(a_0^{0(+)}\to\pi^{0(+)}\eta^{\prime})\simeq 0$.
This leads to 
${\cal B}_{\text{WA}}(\eta)\gg {\cal B}_{\text{WA}}(\eta')\simeq 0$,
which strongly contradicts the relation of
${\cal B}_{\text{WA}}(\eta)\simeq {\cal B}_{\text{WA}}(\eta^{\prime})$.
According to the theoretical studies in Refs.~\cite{Li:2013xsa,Cheng:2019ggx}, 
it is obtained that
${\cal B}_\rho(\eta,\eta^{\prime})=(1.5\pm 0.5,1.2\pm 0.1)\times 10^{-3}$,
which agree with 
${\cal B}_{\rho}(\eta)\simeq {\cal B}_{\rho}(\eta^{\prime})$;
however, with ${\cal B}(\eta)={\cal B}_\rho(\eta)+{\cal B}_{\text{WA}}(\eta)$,
${\cal B}_\rho(\eta)$ leaves tiny room for ${\cal B}(\eta)$ 
to accommodate ${\cal B}_{\text{WA}}(\eta)$.
Therefore, it is reasonable to conclude 
that the $W$-annihilation topologies are unlikely
to be the dominant contributions to $D_{(s)}^+\to\pi^{+(0)}(a_0^{0(+)}\to)\pi^{0(+)}\eta$.

\begin{figure}[t!]
\includegraphics[width=2.8in]{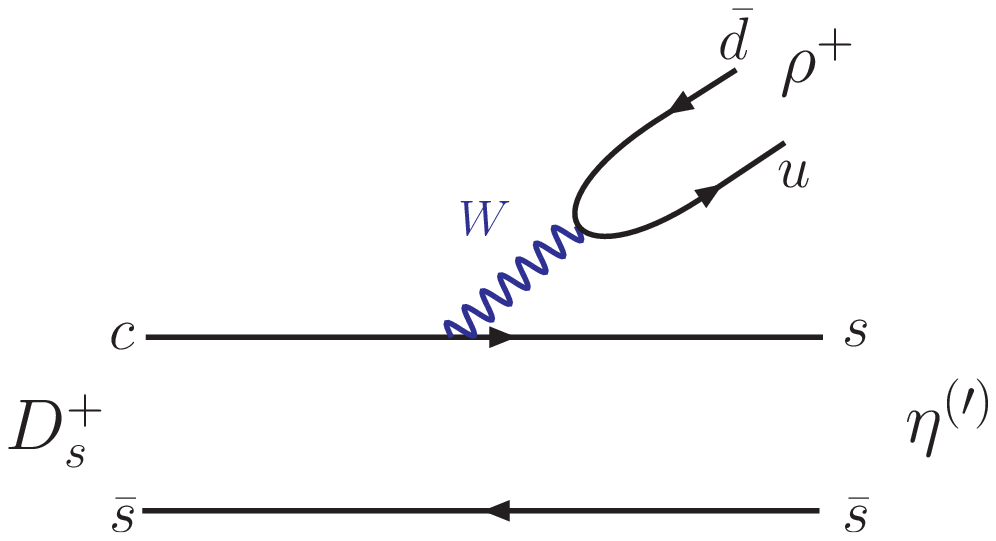}
\caption{The short-distance contribution to $D^+_s\to \eta^{(\prime)}\rho^+$ decay.}\label{Dstoetarho}
\end{figure}

The nearly equal ${\cal B}(D_s^+\to \pi^+ a_0^0,\pi^0 a_0^+)\sim {\it O}(10^{-2})$
are much larger than
the branching fractions of other measured pure $W$-annihilation decays~\cite{Ablikim:2019pit},
such as ${\cal B}(D_s^+\to \pi^+\rho^0)=(2.0\pm 1.2)\times 10^{-4}$.
Besides,
${\cal B}(D_s^+\to \pi^{+(0)} a_0^{0(+)})$ is close to
${\cal B}(D_s^+\to \pi^+ \eta)=(1.70\pm 0.09)\times 10^{-2}$ and
${\cal B}(D_s^+\to \pi^+ f_0(980))\sim {\it O}(10^{-2})$~\cite{pdg},
suggesting that $D_s^+\to \pi^{+(0)}(a_0^{0(+)}\to)\pi^{0(+)}\eta$ 
is more associated with the external $W$-boson emission processes.
Particularly, the $D_s^+\to\eta\rho^+,\eta'\rho^+$ decays proceed through
the external $W$-emission topology as depicted in Fig.~\ref{Dstoetarho}, 
whose branching ratios are observed as large as ${\cal O}(10\%,5\%)$,
respectively~\cite{pdg}.
On the other hand, with ${\cal B}(D_s^+\to \eta\rho)\sim 10\%$,
$D_s^+\to\eta(\rho^+\to)\pi^+\pi^0$ could show up very prominently 
at the $\pi^+\pi^0$ invariant mass spectrum below 1~GeV, which might provide
a possible $\pi^+\pi^0$ final state interaction for the $a_0^+$ formation.
Nonetheless, $\pi^+\pi^0\to a_0^+$ $(a_0^+\to \pi^+\pi^0)$ is a disflavored strong interaction~\cite{pdg}. 
Moreover, without an extra particle emitting to change the helicity state,
the vector to scalar transition through the strong interaction 
should be much suppressed due to the helicity conservation.
We hence propose that,
via the triangle rescattering diagrams in Fig.~\ref{triangle},
$D_s^+\to \pi^{+(0)}(a_0^{0(+)}\to)\pi^{0(+)}\eta$ is able to 
receive the main contributions from $D_s^+\to \eta^{(\prime)}\rho^+$,  
where $\eta^{(\prime)}$ and $\rho^+$ exchange $\pi$ in the final state interaction,
and transform as $a_0$ and $\pi$, respectively.
In this report, we will calculate the $D_s^+\to \pi^{+(0)}(a_0^{0(+)}\to)\pi^{0(+)}\eta$ decays
via the triangle rescattering diagrams, in order to
explain the recent BESIII observation~\cite{Ablikim:2019pit}.\\

\section{Formalism}
%
%
\begin{figure}[t!]
\includegraphics[width=2.8in]{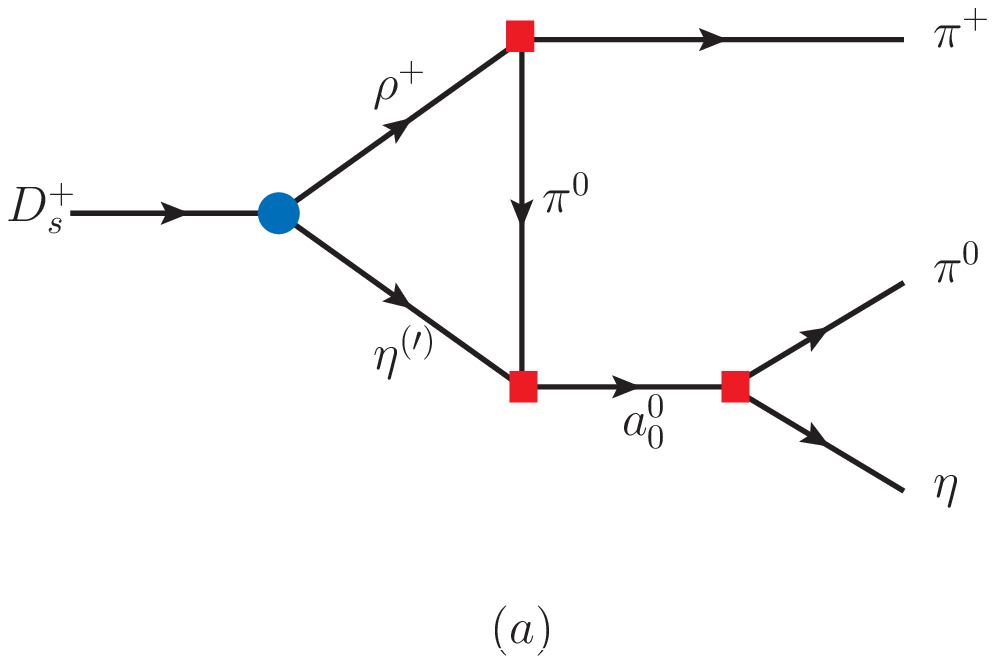}
\includegraphics[width=2.8in]{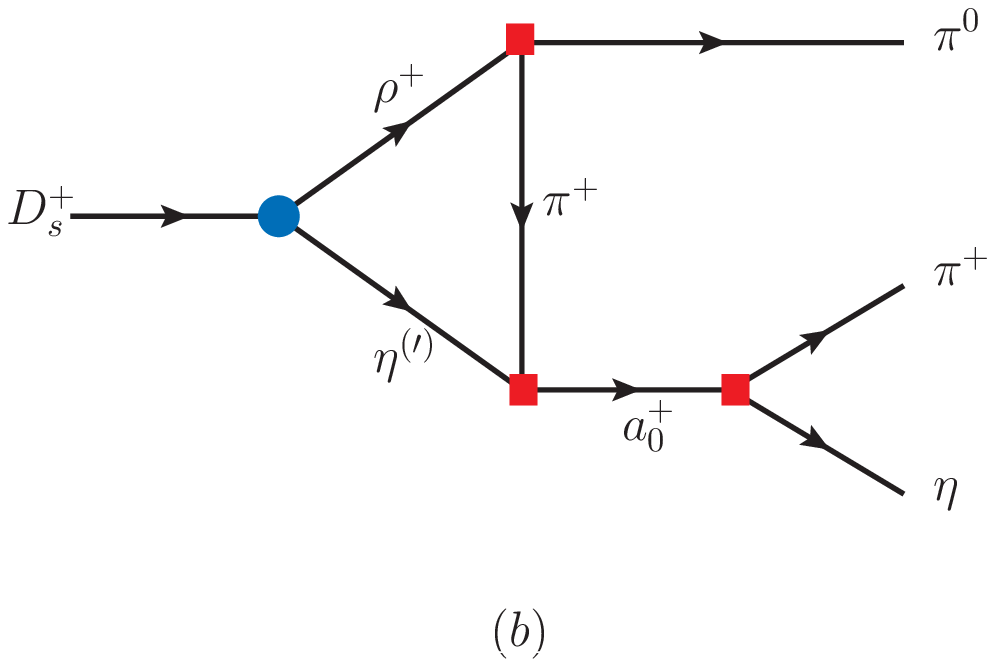}
\caption{The triangle rescattering diagrams for
(a) $D^+_s\to \pi^+(a_0^0\to)\pi^0\eta$ and 
(b) $D_s^+\pi^0(a_0^+\to)\pi^+\eta$.}\label{triangle}
\end{figure}
%
The three-body $D^+_s\to \eta\pi^+\pi^0$ decay predominantly comes from
$D^+_s\to \eta(\rho^+\to)\pi^+\pi^0$. Besides, it receives the contributions from
the $D^+_s\to \pi^+(a_0^0\to)\pi^0\eta$, $\pi^0(a_0^+\to)\pi^+\eta$ decays,
which proceed through the triangle rescattering diagrams in Figs.~\ref{triangle}a and b, respectively.
These resonant $D_s^+$ decays involve $D^+_s\to \eta^{(\prime)}\rho^+$, 
$a_0\to\eta^{(\prime)}\pi$ and $\rho^+\to \pi^+\pi^0$.
For $D^+_s\to \eta^{(\prime)}\rho^+$, the relevant effective Hamiltonian for the $c\to  s u\bar d$ transition 
is given by~\cite{Buras:1998raa}
\begin{eqnarray}\label{Heff}
{\cal H}_{eff}&=&\frac{G_F}{\sqrt 2}
V_{cs}^*V_{ud} [c_1(\bar u d)(\bar s c)+c_2(\bar s d)(\bar u c)]\,,
\end{eqnarray}
where $G_F$ is the Fermi constant, 
$V_{ij}$ the CKM matrix elements,
$c_{1,2}$ the Wilson coefficients, and
$(\bar q_1 q_2)$ stand for $\bar q_1\gamma_\mu(1-\gamma_5)q_2$.
The amplitude of the $D_s^+\to \eta^{(\prime)}\rho^+$ decay 
can be factorized as~\cite{ali}
\begin{eqnarray}\label{amp}
{\cal A}(D_s^+\to \eta^{(\prime)}\rho^+)&=&
\frac{G_F}{\sqrt 2}V_{cs}^*V_{ud} a_1 
\langle \rho^+|(\bar ud)|0\rangle \langle \eta^{(\prime)}|(\bar s c)|D_s^+\rangle\,,
\end{eqnarray}
where $a_1=c_1+c_2/N_c$, with $N_c$ the color number.
The matrix elements in Eq.~(\ref{amp})
are defined by~\cite{Soni:2018adu}
\begin{eqnarray}\label{ffs}
\langle \rho^+|(\bar ud)|0\rangle&=&m_\rho f_\rho \epsilon_\mu^*\,,\nonumber\\
\langle \eta^{(\prime)}|(\bar s c)|D_s^+\rangle&=&
(p_{D_s}+p_{\eta^{(\prime)}})_\mu F_+^{(\prime)}(q^2)+q_\mu F_-^{(\prime)}(q^2)\,,
\end{eqnarray}
with $q_\mu=(p_{D_s}-p_{\eta^{(\prime)}})_\mu$, $\epsilon_\mu^*$ the polarization vector
and $f_\rho$ the decay constant. Besides, the form factor $F_{(\pm)}^{(\prime)}(q^2)$
is in the double-pole parameterization~\cite{Soni:2018adu}:
\begin{eqnarray}\label{t_depent}
F(q^2)=\frac{F(0)}{1-a(q^2/m_{D_s}^2)+b(q^4/m_{D_s}^4)}\,.
\end{eqnarray}
Substituting the matrix elements in Eq.~(\ref{amp}) with those in Eq.~(\ref{ffs}),
we obtain ${\cal A}(D_s^+\to \eta^{(\prime)}\rho^+)$
$=G_{D_{s}\rho\eta^{(\prime)}}\epsilon^*\cdot(p_{D_s}+p_{\eta^{(\prime)}})$
with $G_{D_s \rho\eta^{(\prime)}}\equiv 
(G_F/\sqrt 2)V_{cs}^*V_{ud} a_1 m_\rho f_\rho  F_+^{(\prime)}(m^2_\rho)$,
while $F_-^{(\prime)}(t)$ gives the vanishing contribution due to $\epsilon\cdot q=0$.
For the strong decays $a_0\to\alpha\beta$ and $\rho^+\to \pi^+\pi^0$,
one writes their amplitudes as
\begin{eqnarray}\label{strong}
{\cal A}(a_0\to\alpha\beta)=g_{a_0\alpha\beta},\;\;
{\cal A}(\rho^+\to\pi^+\pi^0)=g_{\rho\pi\pi}\epsilon\cdot (p_{\pi^+}-p_{\pi^0})\,,
\end{eqnarray}
where $\alpha\beta$ could be $\eta^{(\prime)}\pi$ or $K\bar K$, and
$g_{a_0\eta^{(\prime)}\pi}$ and $g_{\rho\pi\pi}$ are the strong coupling constants.
We hence present the amplitudes of the resonant $D_s^+\to \pi^+\pi^0\eta$ decays 
as~\cite{Guo:2017jvc,Li:1996yn,Liu:2017vsf,Liu:2019dqc,Cheng:2004ru,Du:2019idk}
\begin{eqnarray}\label{Are}
{\cal A}_\rho
&\equiv& {\cal A}(D_s^+\to\eta(\rho^+\to)\pi^+\pi^0)
=\frac{-i}{D_1}\hat{\cal A}_\rho(s-t)\,,\; \nonumber\\
{\cal A}_{a(b)}^{(\prime)}
&\equiv& {\cal A}(D_s^+\to \pi^{+(0)}(a_0^{0(+)}\to)\pi^{0(+)}\eta^{(\prime)})
=\frac{1}{D_0}\hat{\cal A}^{(\prime)}{\cal T}_{a(b)}^{(\prime)}\,,\nonumber\\
{\cal T}_{a(b)}^{(\prime)} 
&=&-i\int\frac{d^4 q}{(2\pi)^4 }
\frac{(2p_{D_s}-q)_{\mu}(-g^{\mu\nu}+\frac{q^{\mu}q^{\nu}}{q^2})(q-2p_{\pi^{+(0)}})_{\nu}}
{D_1 D_2 D_3}\,,
\end{eqnarray}
with $s\equiv (p_{\pi^0}+p_{\eta})^2$ and $t\equiv (p_{\pi^+}+p_{\eta})^2$.
Besides, we present that 
$\hat{\cal A}_\rho=G_{D_s\rho\eta}g_{\rho\pi\pi}$,
$\hat{\cal A}=G_{D_s\rho\eta} g^2_{a_0\eta\pi}g_{\rho\pi\pi}$ and
$\hat{\cal A}^{\prime}=G_{D_s\rho\eta^{\prime}} g_{a_0\eta^{\prime}\pi} g_{a_0\eta\pi} g_{\rho\pi\pi}$.
For the propagators in Eq.~(\ref{Are}), 
$D_i$ are given by
\begin{eqnarray}
&&
D_0=x-m_{a_0}^{2}-\sum\limits_{\alpha\beta}
{[\text{Re}\Pi^{\alpha\beta}_{a_0}(m^2_{a_0})-\Pi^{\alpha\beta}_{a_0}(x)]}\,,\nonumber\\
&&D_1=q^2-m_{\rho}^2+im_\rho\Gamma_\rho\,,\nonumber\\
&&D_2=(p_{\pi^{0(+)}}-q)^2-m_{\pi^{0(+)}}^2+i\epsilon^{+}\,,
\nonumber\\
&&D_3=(q-p_{\eta^{(\prime)}})^2-m_{\eta^{(\prime)}}^2+i\epsilon^{+}\,,
\end{eqnarray}
where $x=(s,t)$ for $a_0^{(0,+)}$ in ${\cal A}_{(a,b)}^{(\prime)}$. 
The function of $\Pi^{\alpha\beta}_{a_0}(x)$ in $1/D_0$
is adopted as~\cite{Achasov:2004uq},
\begin{eqnarray}
\Pi^{\alpha\beta}_{a_0}(x)&=& 
\frac{g^{2}_{a_0 \alpha\beta}}{16\pi} 
\bigg\{\frac{m_{\alpha\beta}^+ m_{\alpha\beta}^-}{\pi x}
\log\bigg[\frac{m_\beta}{m_\alpha}\bigg]-\theta[x-(m_{\alpha\beta}^+)^2]\nonumber\\
&\times&
\rho_{\alpha\beta}\bigg(i+\frac{1}{\pi}
\log\bigg[\frac{\sqrt{x-(m_{\alpha\beta}^+)^{2}}+\sqrt{x-(m_{\alpha\beta}^-)^{2}}}
{\sqrt{x-(m_{\alpha\beta}^-)^{2}}-\sqrt{x-(m_{\alpha\beta}^+)^{2}}}\bigg]\bigg)
\nonumber\\
&-&\rho_{\alpha\beta}\bigg(1-\frac{2}{\pi}
\arctan\bigg[\frac{\sqrt{-x+(m_{\alpha\beta}^+)^{2}}}{\sqrt{x-(m_{\alpha\beta}^-)^{2}}}\bigg]\bigg)
(\theta[x-(m_{\alpha\beta}^-)^{2}]-\theta[x-(m_{\alpha\beta}^+)^{2}])\nonumber\\
&+&\rho_{\alpha\beta}\frac{1}{\pi}
\log\bigg[\frac{\sqrt{(m_{\alpha\beta}^+)^{2}-x}+\sqrt{(m_{\alpha\beta}^-)^2-x}}
{\sqrt{(m_{\alpha\beta}^-)^{2}-x}-\sqrt{(m_{\alpha\beta}^+)^2-x}}\bigg]
\theta[(m_{\alpha\beta}^-)^{2}-x]\bigg\}\,,
\end{eqnarray}
where $m_{\alpha\beta}^\pm=m_\alpha\pm m_\beta$ and
$\rho_{\alpha\beta}\equiv \left|\sqrt{x-(m_{\alpha\beta}^+)^{2}}\sqrt{x-(m_{\alpha\beta}^-)^{2}}\right|/x$.
Using $1/D_0$ that presents the propagator of $a_0$,
instead of the Breit-Wigner function like $1/D_1$, we take into account
the contributions from the virtual intermediate states
of $\eta^{(\prime)}\pi$ and $\bar{K}K$,
such that the cusp effect at the threshold of $(m_K+m_{\bar K})$ 
can be given in the $\eta\pi$ invariant mass spectra~\cite{Achasov:2004uq,Bugg:2008ig}.
To proceed, we reduce ${\cal T}_{a,b}$ in Eq.~(\ref{Are}) as
${\cal T}_a={\cal T}(s)$ and ${\cal T}_b=-{\cal T}(t)$,
with ${\cal T}(x)$ given by~\cite{Du:2019idk}
\begin{eqnarray}\label{Tx}
{\cal T}(x)
&=&-i(m_{D_s}^2-m_\rho^2+im_\rho\Gamma_\rho+2m^{2}_{\pi}-2x+m_{\eta}^{2})\int\frac{d^4q}{(2\pi)^4}\frac{1}{D_1D_2D_3}\nonumber\\
&-&i(1+\frac{m_{D_s}^2-m_{\eta}^{2}}{m_\rho^2-im_\rho\Gamma_\rho})\int\frac{d^4q}{(2\pi)^4}\frac{1}{D_1D_3}-i\int\frac{d^4q}{(2\pi)^4}\frac{1}{D_1D_2}\nonumber\\
&+&i\int\frac{d^4q}{(2\pi)^4}\frac{1}{q^2D_1}+i\frac{m_{D_s}^2-m_{\eta}^{2}}{m_\rho^2-im_\rho\Gamma_\rho}
\int\frac{d^4q}{(2\pi)^4}\frac{1}{q^2D_3}+i\int\frac{d^4q}{(2\pi)^4}\frac{1}{D_2D_3}\,,
\end{eqnarray}
where $m_{\pi^0(K^0)}^2\simeq m_{\pi^+(K^+)}^2$ has been used. In the above,
the integrations of the multi-point functions can be found in~\cite{tHooft:1978jhc}.
It is interesting to note that 
the ultraviolet divergences caused by the individual integrations
cancel out~\cite{Du:2019idk,Achasov:2015uua},
such that a cut-off needs not to be introduced in our calculation.
In the same way, we obtain ${\cal T}^{\prime}_{a(b)}$ 
by replacing $\eta$ in ${\cal T}_{a(b)}$ with $\eta'$.
To integrate over the phase space in the three-body decay,
we refer the general equation of the decay width in the PDG~\cite{pdg} 
\begin{eqnarray}\label{gamma1}
\Gamma=\int_{s}\int_{t}
\frac{1}{(2\pi)^3}\frac{|{\cal A}|^2}{32m^3_{D_s}}ds dt\,.
\end{eqnarray}

\section{Numerical Results and Discussions}
\begin{figure}[t!]
\includegraphics[width=2.5in]{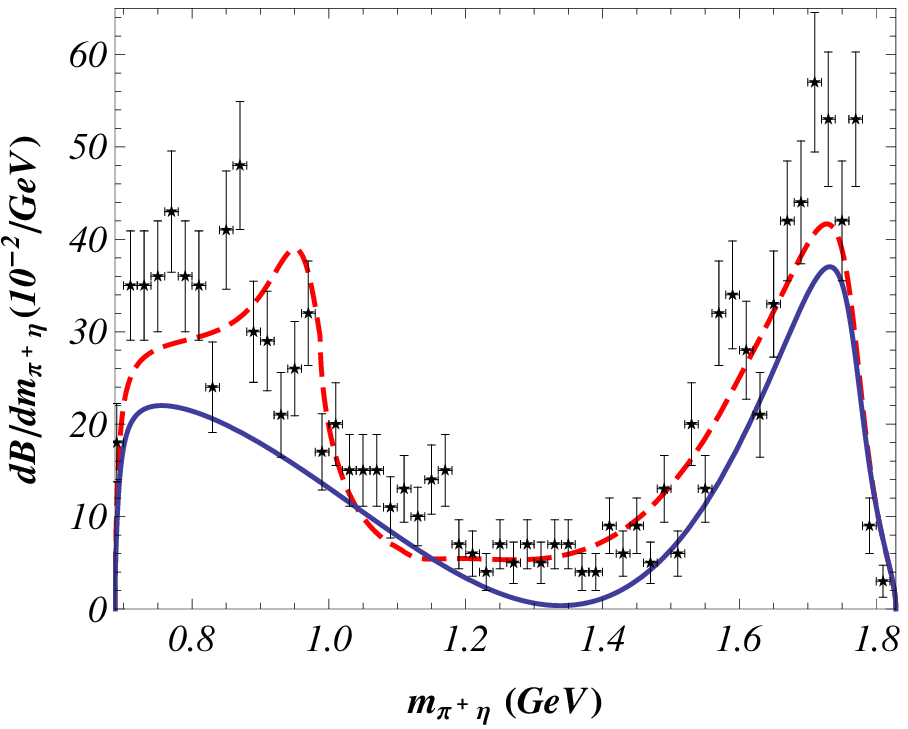}
\includegraphics[width=2.5in]{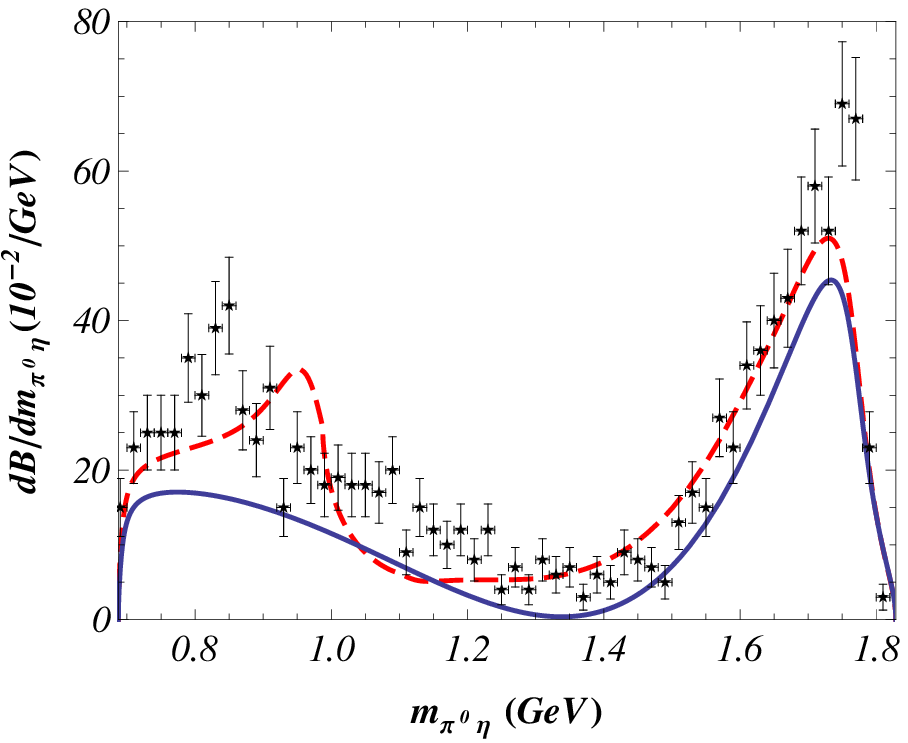}
\caption{The partial distributions vs. $m_{\pi\eta}$, 
where the solid line is for ${\cal A}_\rho$ only, 
while the dashed line receive the contributions from ${\cal A}_\rho$ and ${\cal A}_{a,b}^{(\prime)}$, 
in comparison with the data points in~\cite{Ablikim:2019pit}.}\label{spec1}
\end{figure}

\begin{figure}[t!]
\includegraphics[width=2.5in]{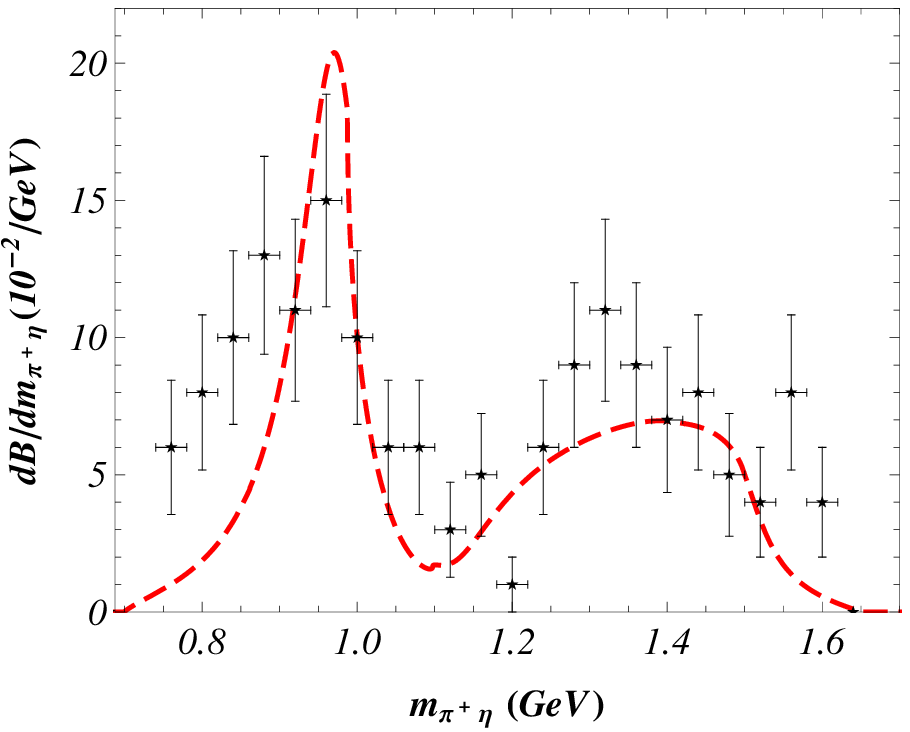}
\includegraphics[width=2.5in]{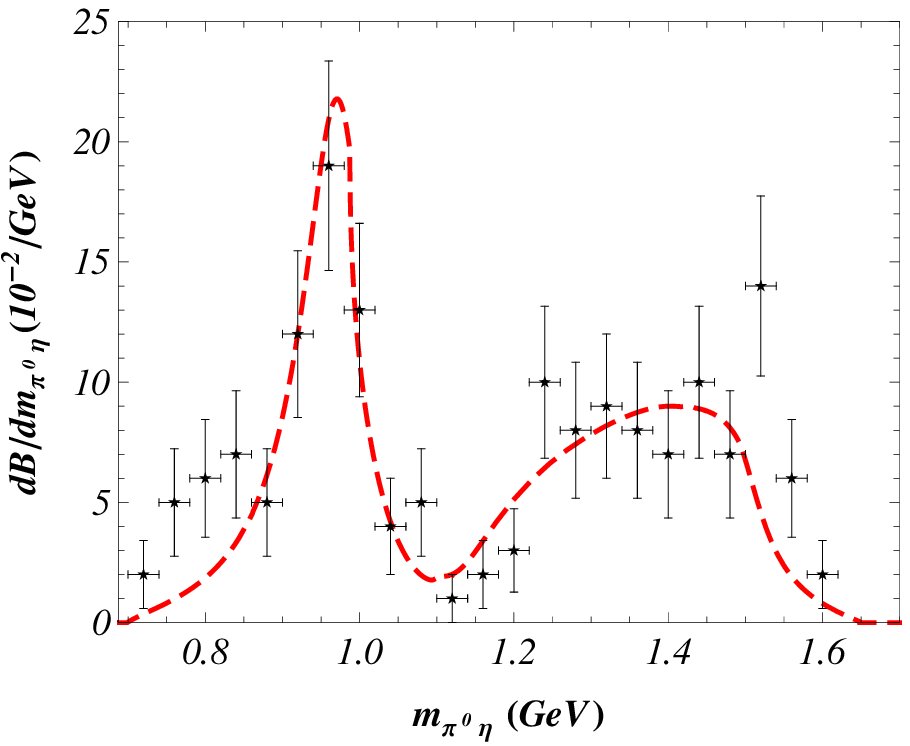}
\caption{The partial distributions vs. $m_{\pi\eta}$ with the cut of $m_{\pi^+\pi^0}>1.0$~GeV, 
in comparison with the data points in~\cite{Ablikim:2019pit}.}\label{spec2}
\end{figure}

In the numerical analysis, we use $V_{cs}=V_{ud}=1-\lambda^2/2$
with $\lambda=0.22453\pm 0.00044$ in the Wolfenstein parameterization
and the decay constant $f_\rho=(210.6\pm 0.4)$~MeV~\cite{pdg}.
For the strong coupling constants, it is given that
$(g_{a_0\eta\pi},g_{a_0\eta'\pi},g_{a_0 K\bar K})=
(2.87\pm 0.09,-2.52\pm 0.08,2.94\pm 0.13)$~GeV~\cite{Bugg:2008ig,Du:2019idk},
while $g_{\rho\pi\pi}=6.0$ is extracted from 
${\cal B}(\rho^+\to \pi^+\pi^0)\simeq100\%$~\cite{pdg}.
We adopt $F_+^{(\prime)}(q^2)$ from Ref.~\cite{Soni:2018adu} as
$(F_+(0),\,a,\,b)=(0.78,\,0.69,\,0.002)$ and 
$(F'_+(0),\,a,\,b)=(0.73,\,0.88,\,0.018)$.
By relating the calculated branching fraction of $D_s^+\to \eta\rho^+$
to the measured value of $(7.4 \pm 0.6)\%$~\cite{Ablikim:2019pit}, 
we determine $a_1=0.93\pm 0.04$, 
where $a_1$ of ${\cal O}(1.0)$ demonstrates 
the validity of the generalized factorization~\cite{ali}.
Consequently, we obtain the branching fractions for 
$D_s^+\to a_0^{+(0)}\pi^{0(+)}$ and $\pi^{+(0)}(a_0^{0(+)}\to)\pi^{0(+)}\eta$ decays, 
\begin{eqnarray}\label{result}
&&{\cal B}(D_s^+\to a_0^{0(+)}\pi^{+(0)})=(1.7\pm 0.2\pm 0.1)\times 10^{-2}\,,\nonumber\\
&&{\cal B}(D_s^+\to \pi^{+(0)}(a_0^{0(+)}\to)\pi^{0(+)}\eta)=(1.4\pm 0.1\pm 0.1)\times 10^{-2}\,,
\end{eqnarray}
where the uncertainties consider the main contributions 
from $a_1$ and $g_{a_0\alpha\beta}$, in order.
We also draw the partial distributions in Figs.~\ref{spec1} and \ref{spec2}
to compare with the data.
%

Our results of the branching fractions, Eq.~(\ref{result}),
agree with the data, Eq.~(\ref{data1}). 
Besides, we predict
${\cal B}(D_s^+\to a_0^0\pi^+)={\cal B}(D_s^+\to a_0^+\pi^0)$,
which agrees with 
the observation that these two-body decays have equal sizes.
The $D_s^+\to \rho\eta$ and $D_s^+\to \rho^+\eta'$ decays 
both give triangle rescattering effects. 
Despite the fact that ${\cal B}(D_s^+\to\rho^+\eta')$ 
is a few times smaller than ${\cal B}(D_s^+\to\rho^+\eta)$~\cite{pdg},
they give similar contributions to ${\cal B}(D_s^+\to a_0\pi)$ and ${\cal B}(D_s^+\to \pi(a_0\to)\eta\pi)$.
Since $\Gamma_\rho\gg \Gamma_{\eta^{(\prime)},\pi}$,
the $\rho$ meson decay width is not negligible,
which causes the width effect~\cite{Achasov:2015uua,Du:2019idk,Guo:2019twa}.
As a test,
we also treat the $\rho$ meson as a stable particle.
Without considering the $\rho$-meson decay width, it is found that
the branching fractions of $D_s^+\to \pi a_0$ and $D_s^+\to \pi(a_0\to)\eta\pi$
are increased by 10\%.

The contributions from $D_s^+\to \pi^+(a_0^0\to)\pi^0\eta$ and 
$D_s^+\to \pi^0(a_0^+\to)\pi^+\eta$ are concluded to
interfere with a relative phase of $180^\circ$ in Ref.~\cite{Ablikim:2019pit}.
With $\rho^+(q)\to\pi^0(q-p_2)\pi^+(p_2)$ and 
$\rho^+(q)\to\pi^+(q-p_2)\pi^0(p_2)$ for ${\cal A}_{a,b}$, respectively,
where $p_2$ is the energy flow for the out-going $\pi$ in the integration,
it leads to ${\cal A}_{a}(\rho^+\to\pi^{+}\pi^{0})=-{\cal A}_{b}(\rho^+\to\pi^{0}\pi^{+})$
from Eq.~(\ref{strong}). Clearly,
the minus sign gives the theoretical explanation to the phase 
of $180^\circ$ in the data.
The $\pi\eta$ invariant mass spectra
in Figs.~\ref{spec1} and \ref{spec2} are demonstrated 
to be consistent with the data~\cite{Ablikim:2019pit}.

\section{Conclusions}
In summary, 
we have proposed that 
$D_s^+\to \pi^{+(0)}(a_0^{0(+)}\to)\pi^{0(+)}\eta$ mainly
proceeds through the triangle loops. By exchanging $\pi^{+(0)}$,
$M^0$ and $\rho^+$ in $D_s^+\to M^0\rho^+$
are formed as $a_0$ and $\pi^{0(+)}$, respectively,
where $M^0=(\eta,\eta')$.
Particularly, we have presented that
${\cal B}(D_s^+\to a_0^{0(+)}\pi^{+(0)})=(1.7\pm 0.2\pm 0.1)\times 10^{-2}$ and
${\cal B}(D_s^+\to \pi^{+(0)}(a_0^{0(+)}\to)\pi^{0(+)}\eta)=(1.4\pm 0.1\pm 0.1)\times 10^{-2}$,
in good agreement with the data.

\section*{ACKNOWLEDGMENTS}
The authors would like to thank Prof.~Liaoyuan Dong for useful discussions.
This work was supported in part by National Science Foundation of China (11675030),
(11905023), and (11875054).


\begin{thebibliography}{99}
\bibitem{Jaffe:2004ph}
R.L.~Jaffe, 
Phys.\ Rept.\  {\bf 409}, 1 (2005). 

\bibitem{Stone:2013eaa} 
S.~Stone and L.~Zhang,
Phys.\ Rev.\ Lett.\  {\bf 111}, 062001 (2013). 

\bibitem{Maiani:2004uc} 
L.~Maiani, F.~Piccinini, A.D.~Polosa and V.~Riquer,
Phys.\ Rev.\ Lett.\  {\bf 93}, 212002 (2004). 

\bibitem{Agaev:2018fvz} 
S.S.~Agaev, K.~Azizi and H.~Sundu,
Phys.\ Lett.\ B {\bf 789}, 405 (2019). 

\bibitem{Wang:2009azc} 
W.~Wang and C.D.~Lu,
Phys.\ Rev.\ D {\bf 82}, 034016 (2010). 

\bibitem{Molina:2019udw} 
R.~Molina, J.J.~Xie, W.H.~Liang, L.S.~Geng and E.~Oset,
Phys.\ Lett.\ B {\bf 803}, 135279 (2020).


\bibitem{Ablikim:2019pit}
M.~Ablikim {\it et al.} [BESIII Collaboration],
Phys.\ Rev.\ Lett.\  {\bf 123}, 112001 (2019). 

\bibitem{Cheng:2010ry} 
H.Y.~Cheng and C.W.~Chiang,
Phys.\ Rev.\ D {\bf 81}, 074021 (2010). 


\bibitem{Achasov:2017edm} 
N.N.~Achasov and G.N.~Shestakov,
Phys.\ Rev.\ D {\bf 96}, 036013 (2017). 

\bibitem{pdg}
M.~Tanabashi {\it et al.} [Particle Data Group],
Phys.\ Rev.\ D {\bf 98}, 030001 (2018).

\bibitem{Li:2013xsa} 
H.n.~Li, C.D.~Lu, Q.~Qin and F.S.~Yu,
Phys.\ Rev.\ D {\bf 89}, 054006 (2014). 

\bibitem{Cheng:2019ggx} 
H.Y.~Cheng and C.W.~Chiang,
Phys.\ Rev.\ D {\bf 100}, 093002 (2019). 

\bibitem{Buras:1998raa}
A.J.~Buras, 
hep-ph/9806471.

\bibitem{ali}
A. Ali, G. Kramer, and C.D. Lu, Phys. Rev.  {\bf D 58}, 094009 (1998).

\bibitem{Soni:2018adu} 
N.R.~Soni, M.A.~Ivanov, J.G.~Korner, 
J.N.~Pandya, P.~Santorelli and C.T.~Tran,
Phys.\ Rev.\ D {\bf 98}, 114031 (2018). 

\bibitem{Li:1996yn} 
X.Q.~Li, D.V.~Bugg and B.S.~Zou,
Phys.\ Rev.\ D {\bf 55}, 1421 (1997).

\bibitem{Cheng:2004ru} 
H.Y.~Cheng, C.K.~Chua and A.~Soni,
Phys.\ Rev.\ D {\bf 71}, 014030 (2005). 

\bibitem{Liu:2017vsf} 
X.H.~Liu and U.G.~Mei$\beta$ner,
Eur.\ Phys.\ J.\ C {\bf 77}, 816 (2017). 

\bibitem{Guo:2017jvc} 
F.K.~Guo, C.~Hanhart, U.G.~Mei${\beta}$ner, Q.~Wang, Q.~Zhao and B.S.~Zou,
Rev.\ Mod.\ Phys.\  {\bf 90}, 015004 (2018). 

\bibitem{Liu:2019dqc} 
X.H.~Liu, G.~Li, J.J.~Xie and Q.~Zhao,
Phys.\ Rev.\ D {\bf 100}, 054006 (2019). 

\bibitem{Du:2019idk} 
M.C.~Du and Q.~Zhao,
Phys.\ Rev.\ D {\bf 100}, 036005 (2019). 
 
\bibitem{Achasov:2004uq} 
N.N.~Achasov and A.V.~Kiselev,
Phys.\ Rev.\ D {\bf 70}, 111901 (2004). 

\bibitem{Bugg:2008ig} 
D.V.~Bugg, 
Phys.\ Rev.\ D {\bf 78}, 074023 (2008). 

\bibitem{tHooft:1978jhc} 
G.~$'$t~Hooft and M.J.G.~Veltman,
Nucl.\ Phys.\ B {\bf 153}, 365 (1979).

\bibitem{Achasov:2015uua} 
N.N.~Achasov, A.A.~Kozhevnikov and G.N.~Shestakov,
Phys.\ Rev.\ D {\bf 92}, 036003 (2015). 

\bibitem{Guo:2019twa} 
F.K.~Guo, X.H.~Liu and S.~Sakai,
Prog.\ Part.\ Nucl.\ Phys.\  {\bf 112}, 103757 (2020). 

\end{thebibliography}
\end{document}